\documentclass{PoS}
\usepackage{amsmath}
\usepackage{subfigure} 
\usepackage{wrapfig}
\allowdisplaybreaks[1]
\newcommand{\e}{{\rm{e}}}
\newcommand{\td}{{\rm{d}}}

\title{Non-perturbative renormalization of bilinear operators with M\"{o}bius domain-wall fermions\\
in the coordinate space}

\ShortTitle{NPR of bilinears with M\"{o}bius domain-wall fermions in the coordinate space}

\author{JLQCD Collaboration: \speaker{M. Tomii}$^{a,b}$,%\thanks{A footnote may follow.}\\
	\ G. Cossu$^b$, S. Hashimoto$^{a,b}$, J. Noaki$^b$\\
        E-mail: \email{tomii@post.kek.jp}
        	}

\author{\it $^a$ School of High Energy Accelerator Science, The Graduate University for Advanced Studies
(Sokendai), Ibaraki 305-0801, Japan,}
\author{\it $^b$ KEK Theory Center, Institute of Particle and Nuclear Studies, High Energy Accelerator Research Organization (KEK), Ibaraki 305-0801, Japan}
\abstract{
We study the non-perturbative determination of the renormalization constants of flavor non-singlet quark bilinear operators on the lattice. The renormalization condition is imposed on correlation functions of bilinear operators in the coordinate space. The results are converted to the value at 2 GeV in the $\rm\overline{MS}$ scheme by a perturbative matching. The calculation is carried out on gauge configurations generated with the Mobius domain-wall fermions at two lattice spacings $a^{-1} = 2.4$ GeV and $a^{-1} = 3.6$ GeV.
}

\FullConference{The 32nd International Symposium on Lattice Field Theory,\\
		23-28 June, 2014\\
		Columbia University New York, NY}

\begin{document}

\section{Introduction}

There are several methods to renormalize the lattice operators onto those defined in the
continuum renormalization scheme such as the $\rm\overline{MS}$ scheme.
Schr\"{o}dinger functional \cite{Luscher_etal:1992} and RI/MOM \cite{Martinelli_etal:1995}
approaches are among the popular choices.
The former provides precise and fully non-perturbative renormalization at the cost of
generating dedicated ensembles.
The renormalization condition in the latter approach is imposed on
the vertex function at a fixed gauge and it involves (continuum) perturbative expansion
which is calculated only at one- or two-loop order.

In this work, we investigate the determination of the renormalization constants by the X-space method,
which was originally suggested in \cite{Martinelli_etal:1997} and
has been developed in \cite{V.Gimenez_etal:2004,K.Cichy_etal:2012}.
In this method, the renormalization condition is imposed on two-point
correlation functions of the operators to be renormalized.
Unlike RI/MOM, the X-space method enables us to renormalize by a gauge invariant
quantity.
By keeping the distance finite, one can avoid extra divergences due to contact terms.
Another important advantage is that the perturbative expansion on the continuum side
is available to the four-loop level for the quark bilinear operators \cite{Chetyrkin_Maier_2011}.
On the other hand, a potential problem of the X-space method is the requirement for a
"window" where the continuum perturbation theory can be applied and the lattice
calculation with minimal discretization effect is possible.

In this report, we present the preliminary results of the determination of renormalization
constants of
flavor non-singlet quark bilinear operators using the X-space method.
We obtain good precision on the lattices generated with
the $2+1$ flavors of M\"{o}bius domain-wall fermions
and Symanzik improved gauge action \cite{Noaki_etal:2014}.
We work on $32^3\times64$ lattices at $a^{-1} = 2.4$ GeV and $48^3\times96$
lattices at $a^{-1}=3.6$ GeV, both having matched physical volume and similar input
physical masses in the window of $M_\pi : 300 \sim 500$ MeV.

\section{Sketch of the X-space method}

We impose the renormalization condition on massless correlation functions of
(flavor non-singlet) quark bilinear operators, which are defined as
\begin{align}
&\Pi_{\rm SS}(x) = \big\langle S(x)S(0)\big\rangle,\ \ \ \ 
\Pi_{\rm PP}(x) = \big\langle P(x)P(0)\big\rangle,\\
&\Pi_{\rm VV}(x) = \sum_{\mu=1}^4\big\langle V_\mu(x)V_\mu(0)\big\rangle,\ \ \ \ 
\Pi_{\rm AA}(x) = \sum_{\mu=1}^4\big\langle A_\mu(x)A_\mu(0)\big\rangle,
\end{align}
where $S$ and $P$ are scalar and pseudoscalar densities, $V_\mu$ and $A_\mu$ are
vector and axial-vector currents.
Flavor indices are omitted for simplicity, but they are understood as isospin triplet
operators of light quarks.
These bilinear operators on the lattice are to be renormalized onto the
$\rm\overline{MS}$ scheme at 2 GeV, i.e.
\begin{equation}
{O}_\Gamma^{\rm\overline{MS}}|_{2\rm\ GeV}
= Z_\Gamma^{\overline{\rm MS}/lat}(2{\rm\ GeV};a){O}_\Gamma^{lat}|_a,
\end{equation}
for each quark bilinear operators ${O}_\Gamma\in\{S, P, V_\mu, A_\mu\}$.

Since these correlation functions involve two bilinear operators, the renormalization condition
is given by
\begin{equation}
{Z_\Gamma^{\overline{\rm MS}/lat}(2{\rm\ GeV};a)}^2
\Pi_{\Gamma\Gamma}^{lat}(x)|_a = \Pi_{\Gamma\Gamma}^{\rm\overline{MS}}(x)|_{\rm 2\ GeV},
\label{eq:renorm_cond}
\end{equation}
or
\begin{equation}
Z_\Gamma^{\overline{\rm MS}/lat}(2{\rm\ GeV};a)
= \sqrt{\Pi_{\Gamma\Gamma}^{\rm\overline{MS}}(x)|_{\rm 2\ GeV}\over\Pi_{\Gamma\Gamma}^{lat}(x)|_a},
\label{eq:RCs}
\end{equation}
where $\Pi_{\Gamma\Gamma}^{\rm\overline{MS}}$ and $\Pi_{\Gamma\Gamma}^{lat}$
are correlation functions calculated in the continuum theory or measured on the lattice.

We need two calculations to obtain the renormalization constants:
\begin{itemize}
\item $\Pi_{\Gamma\Gamma}^{lat}(x)|_a:$ Correlation functions calculated on the lattice and taken the chiral limit.
\item $\Pi_{\Gamma\Gamma}^{\rm\overline{MS}}(x)|_{\rm 2\ GeV}:$ Massless continuum correlation functions renormalized at 2 GeV in the
$\rm\overline{MS}$ scheme.
\end{itemize}

In this process, we have to choose $x$ in the window $a\ll |x|\ll\Lambda_{\rm QCD}^{-1}$
in order to avoid discretization effect and perturbative ambiguity.

\section{Correlation functions measured on the lattice}

\begin{figure}[tb]
\vspace{-4.2mm}
\begin{tabular}{cc}
\begin{minipage}{0.47\hsize}
\vspace{4.2mm}
\begin{center}
\includegraphics[width=\hsize,clip]{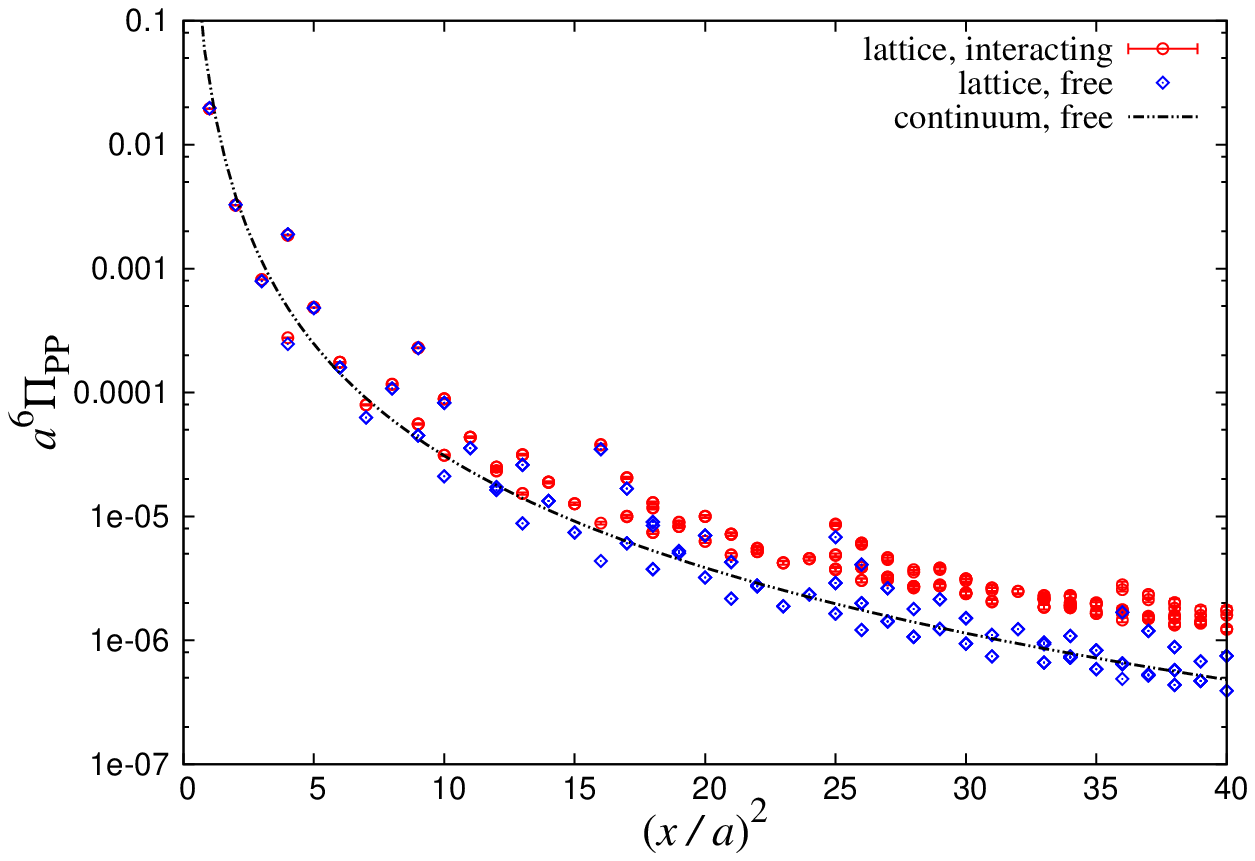}
\caption{Correlation function for P channel which are purely measured on the lattice in the ensemble
$48^3\times96,\ \beta=4.35,\ am_{ud}=0.012,\ am_s=0.018$ and that in the free system
plotted with that in free continuum theory.
}
\label{fig:gatagata}
\end{center}
\end{minipage}
\quad
\begin{minipage}{0.47\hsize}
\begin{center}
\includegraphics[width=\hsize,clip]{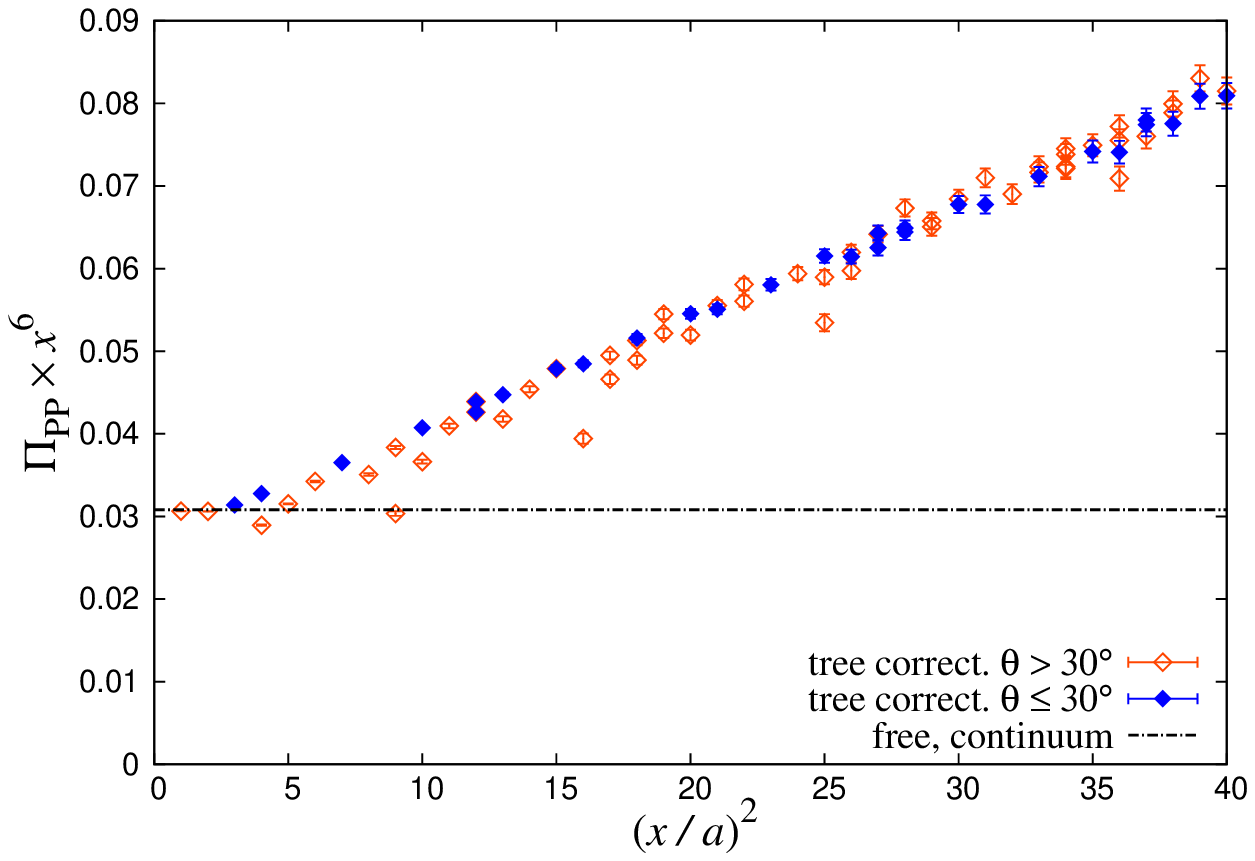}
\caption{
Correlation function after applying tree-level correction to that in Fig.~1 with
the discrimination of the range $\theta > 30^\circ$ (open diamond) from
$\theta\le30^\circ$ (filled diamond).
}
\label{fig:free_correction}
\end{center}
\end{minipage}
\end{tabular}
\end{figure}

Figure~\ref{fig:gatagata} shows $x^2$-dependence of short-distance lattice correlator
of pseudoscalar channel plotted as a function of $x^2$ (red).
In this plot, we also show the two-point correlator calculated at the tree-level,
i.e. no strong interaction (blue).
The data show substantial violation of the rotational symmetry, but apparently
their short-distance behavior is rather precisely reproduced by the free propagator.
The simplest example is the two-points at $(x/a)^2=4$ where $(2,0,0,0)$ and $(1,1,1,1)$
are ten-times different, but they are well reproduced at tree-level.

%=   Figure   ======================
%===============================

We can thus eliminate the bulk of discretization effect by subtracting
the tree-level contribution of the discretization effect as

\begin{equation}
\Pi_{\Gamma\Gamma}^{lat} (x) \longrightarrow \Pi_{\Gamma\Gamma}^{lat} (x)
- \left( \Pi_{\Gamma\Gamma}^{lat,free}(x) - \Pi_{\Gamma\Gamma}^{cont,free}(x) \right).
\label{eq:tree_correction}
\end{equation}
The results are shown in Fig.~\ref{fig:free_correction}.
We note that the vertical axis is in the linear scale rather than logarithmic.
Reductions of the Lorentz violation is quite clear.
A similar subtraction was introduced, but multiplicatively,
i.e. multiplying the original correlator $\Pi_{\Gamma\Gamma}^{lat}(x)$ by the ratio
$\Pi_{\Gamma\Gamma}^{cont,free}(x) / \Pi_{\Gamma\Gamma}^{lat,free}(x)$.
We find that (\ref{eq:tree_correction}) works slightly better to eliminate the Lorentz violating effect.

Another way to reduce discretization effect is suggested \cite{K.Cichy_etal:2012}
on top of (\ref{eq:tree_correction}).
Let us define $\theta$ as an angle between $x$ and the direction (1,1,1,1).
One may find that
the discretization effect becomes more pronounced for large $\theta$.
By inspecting the results, we decided to drop the points of $\theta>30^\circ$.
In Fig.~\ref{fig:free_correction}, the discarded points are shown by open symbols.
The remaining points (blue diamonds) are quite smooth as a function of $(x/a)^2$.

\section{Correlation functions computed from continuum perturbation theory}

%===============================
\begin{figure}
\begin{center}
\subfigure{\mbox{\raisebox{1mm}{\includegraphics[width=73mm]{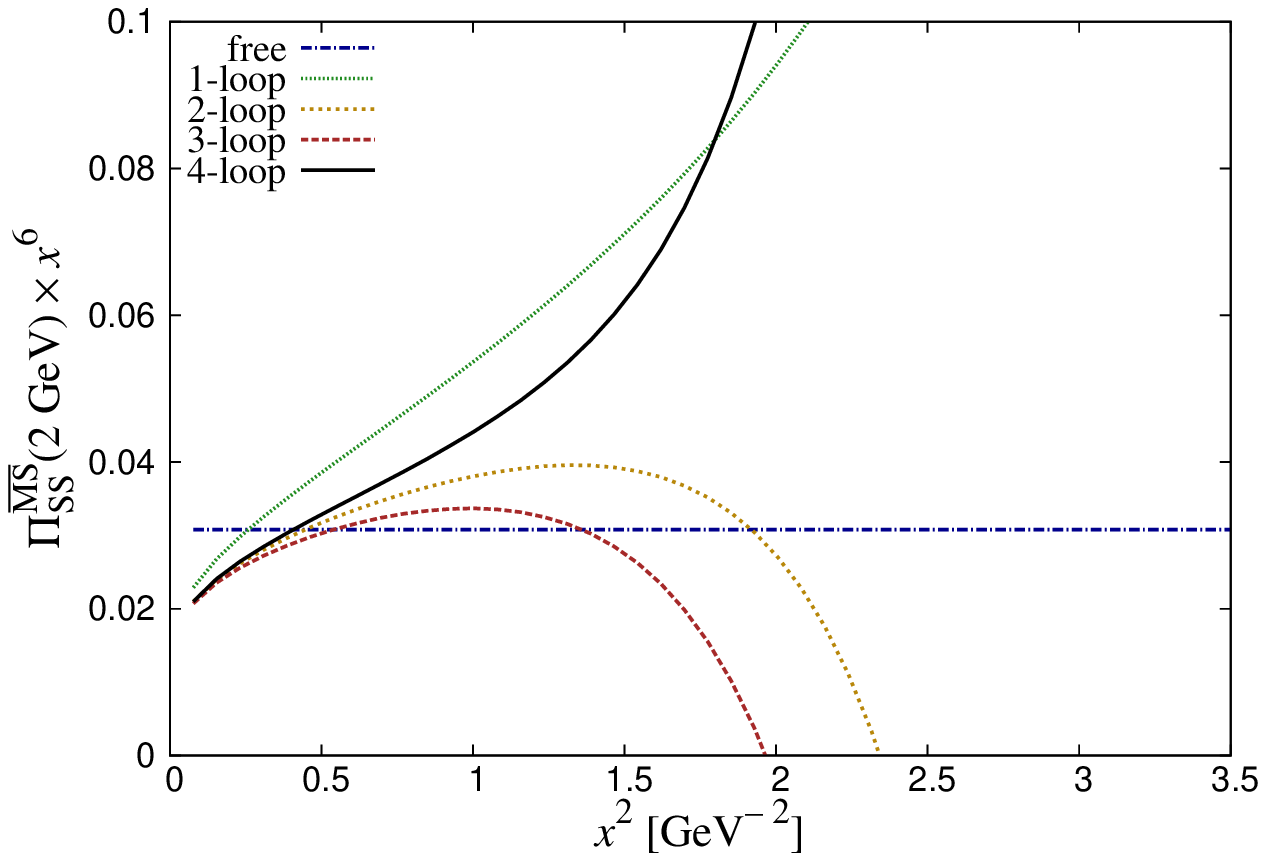}}}}
\subfigure{\mbox{\raisebox{1mm}{\includegraphics[width=73mm]{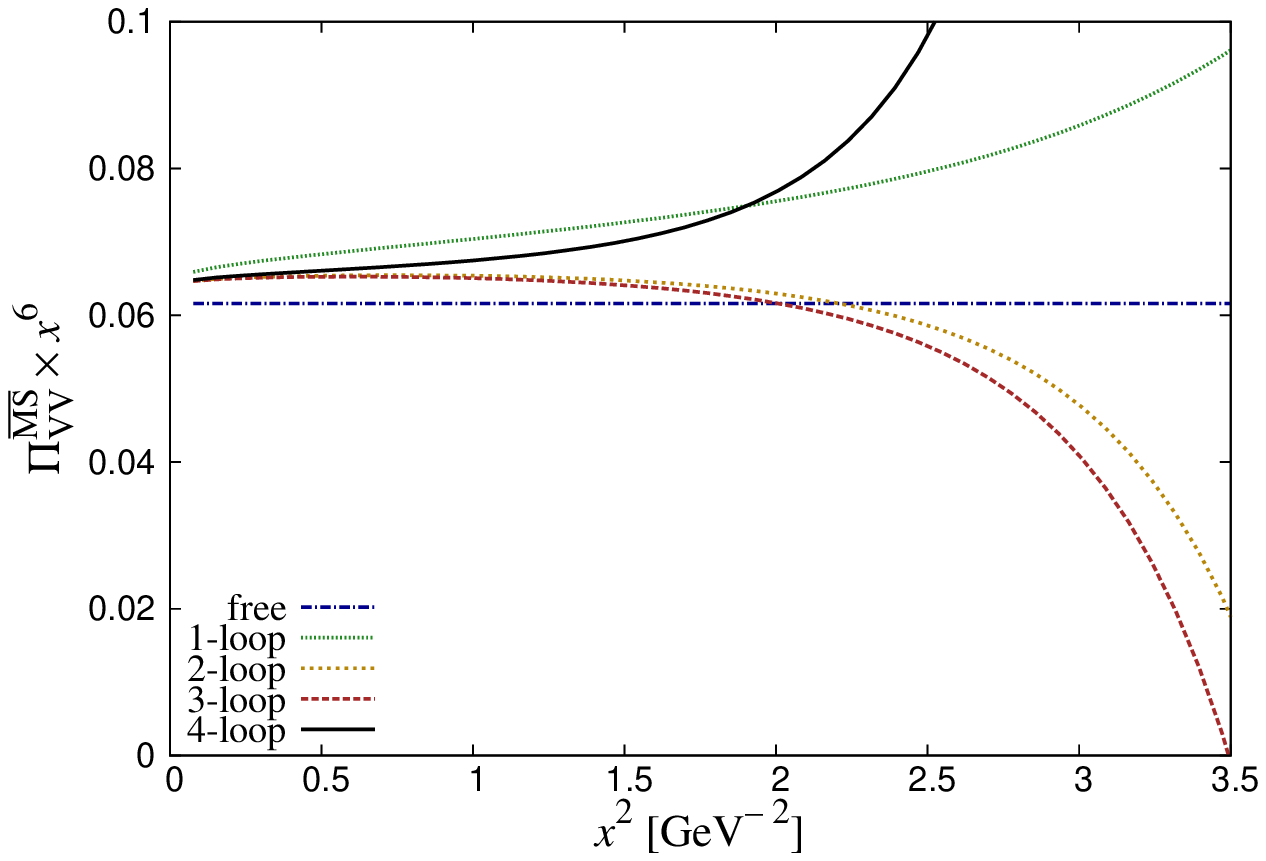}}}}
\caption{
Massless correlators from perturbation theory with $N_f=3,\ \Lambda_{\rm QCD}=340$
MeV, scalar channel at 2 GeV in $\rm\overline{MS}$ scheme (left panel)
and vector channel (right panel).
}
\label{fig:x2vsx6corr_cont}
\end{center}
\end{figure}
%===============================

Perturbative expansion of the massless correlators in the $\rm\overline{MS}$ scheme
\begin{align}
&\Pi_{\rm SS,PP}^{\rm\overline{MS}}(x,\mu) = \Pi_{\rm SS,PP}^{\rm\widetilde{MS}}(x,\tilde\mu)
= {3\over\pi^4x^6}\left(1+\sum_n\widetilde C_n^{\rm S}\tilde a_s^n\right),
\label{eq:PE_scalar}
\\
&\Pi_{\rm VV,AA}^{\rm\overline{MS}}(x) = \Pi_{\rm VV,AA}^{\rm\widetilde{MS}}(x)
= {6\over\pi^4x^6}\left(1+\sum_n\widetilde  C_n^{\rm V}\tilde a_s^n\right),
\label{eq:PE_vector}
\\
&\tilde a_s = {\alpha_s^{\rm\widetilde{MS}}(\tilde\mu = 1/x)\over\pi}
 = {\alpha_s^{\rm\overline{MS}}(\mu = 2\e^{-\gamma_E}/x)\over\pi},
\end{align}
are known up to $n = 4$ \cite{Chetyrkin_Maier_2011}.
$\widetilde C_n^{\rm S, V}$ are perturbative coefficients and $a_s = \alpha_s/\pi$.
The running of $\alpha_s$ is also known up to four-loop level \cite{Chetyrkin_etal_1997}
in the $\rm\overline{MS}$ scheme.

Since the scale $\mu$ and $\tilde\mu$ in these formulae depend on $x$,
we need to perform the scale evolution for correlation functions of scalar and
pseudoscalar correlators to those at 2 GeV in the $\rm\overline{MS}$ scheme.
The scale evolution is calculated as
\begin{equation}
\Pi_{\rm SS, PP}^{\rm\widetilde{MS}}(x,\tilde\mu_1)
= \left[{c(a_s(\tilde\mu_1))\over c(a_s(\tilde\mu_0))}\right]^{-2}
\Pi_{\rm SS, PP}^{\rm\widetilde{MS}}(x,\tilde\mu_0),
\ \ \ \ 
c(x) \equiv \exp\left[\int^x\td x' {\gamma_m(x')\over\beta(x')}\right],
\label{eq:scale_evolve}
\end{equation}
where $\gamma_m$ and $\beta$ are the quark mass anomalous dimension and
the QCD beta function,
and $c(x)$ is given in \cite{K.G.Chetyrkin_1997,J.A.M.Vermaseren_etal:1997}.
The Ward-Takahashi identity guarantees that the correlation functions of vector or
axial-vector currents are scale independent.

%===============================
\begin{figure}
\begin{center}
\subfigure{\mbox{\raisebox{1mm}{\includegraphics[width=73mm]{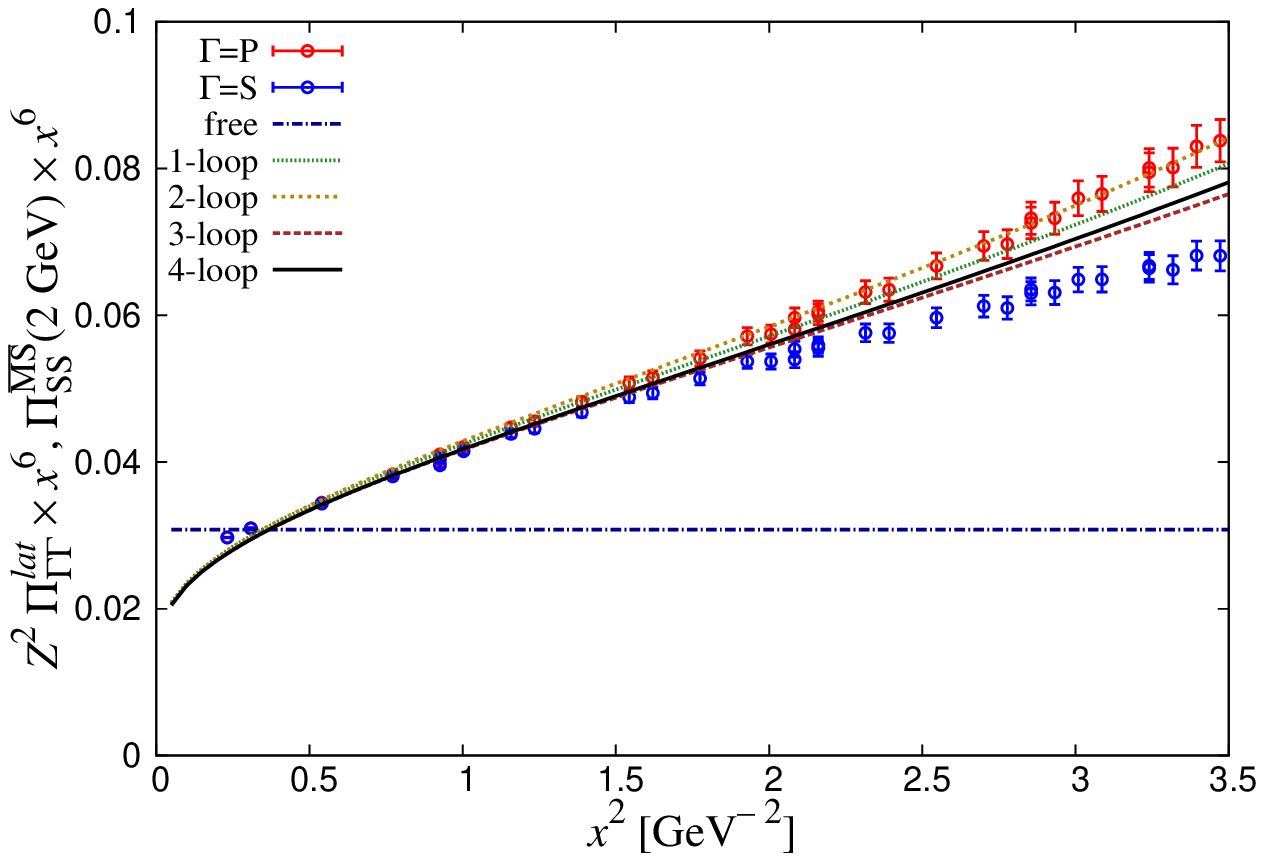}}}}
\subfigure{\mbox{\raisebox{1mm}{\includegraphics[width=73mm]{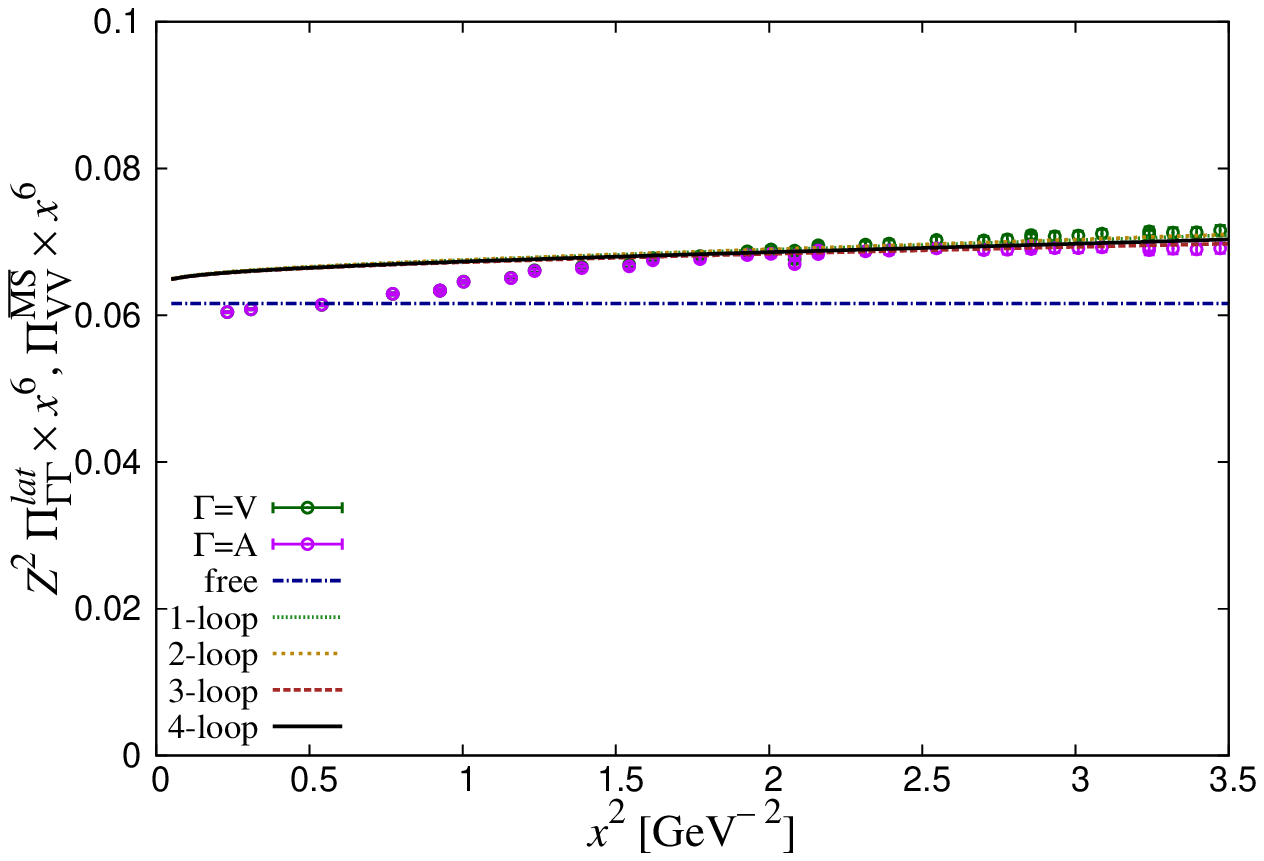}}}}
\caption{
Massless correlators from perturbation theory with improved convergence and
chiral limits of lattice correlators times some constants, which are squares of renormalization
constants.
These lattice correlators are measured on $48^3\times96,\ a^{-1} = 3.6$ GeV lattice,
while a sea quark is still massive $am_s=0.0180$.
}
\label{fig:x2vsx6corr_lat_cont}
\end{center}
\end{figure}
%===============================

As shown in Fig.~\ref{fig:x2vsx6corr_cont}, the convergence of the perturbative series in
(\ref{eq:PE_scalar}) and (\ref{eq:PE_vector}) is not sufficiently good to achieve precise
determination of the renormalization constant already at $x\simeq\rm1\ GeV^{-1}$.
At longer distances, the expansion is even divergent.
The perturbative series in $N_f=3$ theory are given as
\begin{align}
&\Pi_{\rm SS}^{\rm\widetilde{MS}}(x,\tilde\mu)
= {3\over\pi^4x^6}(1 + 0.67\tilde a_s - 16.3\tilde a_s^2 - 31\tilde a_s^3 + 497 \tilde a_s^4),
\\
&\Pi_{\rm VV}^{\rm\widetilde{MS}}(x)
= {6\over\pi^4x^6}(1+\tilde a_s - 4\tilde a_s^2 - 1.9\tilde a_s^3 + 94\tilde a_s^4).
\end{align}
When $\tilde a_s=0.1$, the expansion for SS is like
$1+0.067-0.163-0.031+0.050$ and does not seem to converge.

In order to avoid this problem, we expand correlators in terms of the coupling $a_s^*$
at another scale $\mu^*$ according to the BLM prescription \cite{BLM:1983}.
In this way, one can
effectively absorbs higher-order contribution from vacuum polarization effects into lower-orders.
The scale for VV (and AA) is given by $\mu^* = \mu\e^{-11/6 + 2\zeta(3)} \simeq 1.8\mu$.
We use the same scale also for SS (and PP).

The resulting expansion becomes
\begin{align}
&\Pi_{\rm SS}^{\rm\widetilde{MS}}(x, \mu^*)
= {3\over\pi^4x^6}(1+2.9 a_s^* + 1.1 a_s^{*2} - 42 a_s^{*3} + 24 a_s^{*4}),
\\
&\Pi_{\rm VV}^{\rm\widetilde{MS}}(x)
= {6\over\pi^4x^6}(1+ a_s^* + 0.083 a_s^{*2} - 6 a_s^{*3} + 18 a_s^{*4}),
\end{align}
which obviously has much better convergence property.
The scale evolution for $\Pi_{\rm SS}$ to 2 GeV in $\rm\overline{MS}$
is then calculated using (\ref{eq:scale_evolve}).

Figure~\ref{fig:x2vsx6corr_lat_cont} shows the continuum correlators obtained
with the above procedure
at each order of the perturbative expansion.
It clearly shows a good convergence even at lower scales $x^2\sim3\rm\ GeV^{-2}$.

%============================
%\begin{wrapfigure}[12]{r}{.3\linewidth}
%\begin{wrapfigure}{r}{7cm}
\begin{figure}[t]
\begin{center}
%\vspace{-4.5mm}
\includegraphics[width=10.5cm]{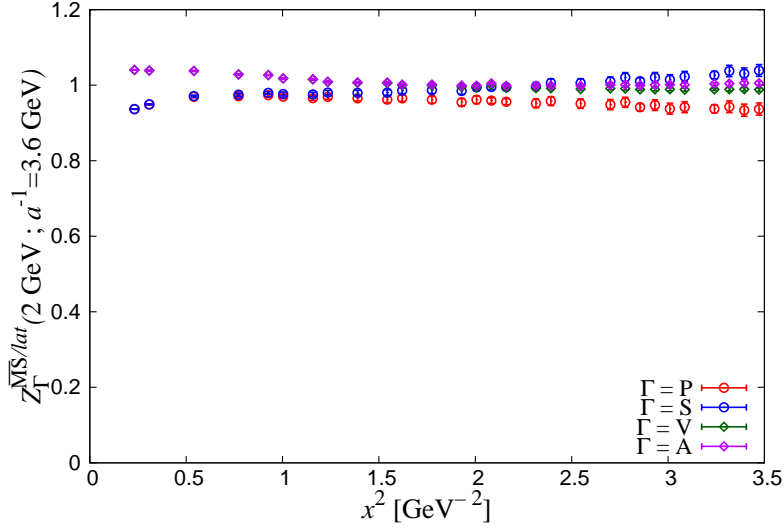}
\caption{
Renormalization constants with the dependence on the renormalization point $x$,
where the parameters are same as Fig.~4.
}
\label{fig:RCs}
%\vspace{-8mm}
\end{center}
\end{figure}
%\end{wrapfigure}
%============================

%===========================
%\begin{wraptable}{r}{8cm}
\begin{table}[h]
\begin{center}
%\vspace{-10mm}
\caption{
Preliminary results of computing renormalization constants. First error means statistical error and second one means
systematic error.}
\vspace{1mm}
\label{tab:PreRst}
\begin{tabular}{cccc}
  \hline
  \hline
$a^{-1}$ [GeV] & $am_s$ & $Z_S^{\overline{\rm MS}/lat}$(2 GeV) & $Z_V^{\overline{\rm MS}/lat}$\\
\hline
%__Start__
$2.4$ & $0.030$ & 1.092(6)(12) & 1.013(2)(7) \\
$2.4$ & $0.040$ & 1.093(8)(28) & 1.017(2)(9) \\
\hline
$3.6$ & $0.018$ & 0.973(5)(6) & 0.999(3)(5) \\
$3.6$ & $0.025$ & 0.965(7)(11) & 0.994(6)(5) \\
%__Last__
\hline
\hline
\end{tabular}
\end{center}
\end{table}
%\end{wraptable}
%===========================

\section{Preliminary results}

Figure~\ref{fig:x2vsx6corr_lat_cont} also shows the lattice data rescaled by a factor
such that they agree with the perturbation theory in the region of $x^2\sim\rm2\ GeV^{-2}$.
This factor corresponds to the renormalization factor squared as given in
(\ref{eq:renorm_cond}).

Figure~\ref{fig:RCs} shows the renormalization constant calculated from
eq.~(\ref{eq:RCs}) with dependence on the renormalization point $x$.
Since LHS of eq.~(\ref{eq:RCs}) is independent of $x$,
we should extract renormalization constants from a range
where $x$-dependence of them is approximately absent.
Since we employ the M\"{o}bius domain wall fermions, which have
excellent chiral symmetry,
we can assume $Z_{\rm S} = Z_{\rm P},\ Z_{\rm V} = Z_{\rm A}$ in the chiral limit, and
extract $Z_{\rm S}$ and $Z_{\rm V}$ as an average of $\{Z_{\rm S}, Z_{\rm P}\}$
or $\{Z_{\rm V}, Z_{\rm A}\}$, respectively.
Here the difference between $Z_{\rm S}$ and $Z_{\rm P}$, or $Z_{\rm V}$ and $Z_{\rm A}$,
coming from higher orders of the operator product expansion is
considered as a part of systematic errors.

Table~\ref{tab:PreRst} shows the preliminary results for the renormalization constants.
Since we calculate only at two strange quark masses for each lattice spacings,
we do not take the chiral limit of $m_s$.
Systematic errors, which are the second errors in Table~\ref{tab:PreRst},
are typically $1\%$ or less except for
$Z_{\rm S}$ at $a^{-1}=2.4$ GeV with $am_s = 0.040$.
This $1\%$ precision for $Z_{\rm S}$
is better than that of other methods, such as RI/MOM at similar lattice spacings.

\section{Summary}

We investigate the non-perturbative renormalization of flavor non-singlet quark
bilinear operators using the gauge invariant X-space method for the action of
$2+1$ M\"{o}bios domain-wall fermions and Symanzik improved gauge action.

Discretization effect of correlation functions measured on the lattice is substantially
reduced by applying the tree-level correction and the angle cut.
Convergence of perturbation theory
is dramatically improved by expanding correlators as
a polynomial of coupling $a_s^*$.
The systematic error we can reach is roughly $1\%$ or better.

\begin{acknowledgments}
Numerical simulations are performed on the IBM System Blue Gene Solution at High Energy
Accelerator Research Organization (KEK) under a support of
its Large Scale Simulation Program
(No. 13/14-04). This work is supported in part by the Grant-in-Aid of the Japanese Ministry of
Education (No. 26247043) and the SPIRE (Strategic Program
for Innovative Research) Field5 project.
\end{acknowledgments}

\end{document}